\documentclass[twocolumn]{aastex631}
\usepackage{graphicx}
\usepackage{placeins}
\usepackage{amsmath}
\usepackage{amssymb}
\usepackage{natbib}
\usepackage{graphicx}
\let\tablenum\relax
\usepackage{xcolor}
\usepackage{rotating}
\usepackage{float}

\newcommand{\z}{$1<z\lesssim2$\;}

\newcommand*\cleartoleftpage{%
  \clearpage
  \ifodd\value{page}\hbox{}\newpage\fi
}

\def \m {MAGPHYS\;}

\def\nar{\ref@jnl{New A Rev.}}          

\shorttitle{Accretion and Jet Coupling in Powerful Radio Quasars at Cosmic Noon}
\shortauthors{Azadi et al.}

\begin{document}

\title{Accretion and Jet Coupling in Powerful Radio Quasars at Cosmic Noon}

\author{Mojegan Azadi}
\affiliation{Center for Astrophysics $|$ Harvard \& Smithsonian, 60 Garden Street, Cambridge, MA, 02138, USA}

\author{Belinda Wilkes}
\affiliation{Center for Astrophysics $|$ Harvard \& Smithsonian, 60 Garden Street, Cambridge, MA, 02138, USA}

\author{Joanna Kuraszkiewicz}
\affiliation{Center for Astrophysics $|$ Harvard \& Smithsonian, 60 Garden Street, Cambridge, MA, 02138, USA}

\author{S.~P. Willner}
\affiliation{Center for Astrophysics $|$ Harvard \& Smithsonian, 60 Garden Street, Cambridge, MA, 02138, USA}

\author{Matthew L.~N. Ashby}
\affiliation{Center for Astrophysics $|$ Harvard \& Smithsonian, 60 Garden Street, Cambridge, MA, 02138, USA}


\begin{abstract}

We present bolometric corrections, as a function of wavelength, for powerful radio-loud quasars from the Revised Third Cambridge Catalogue of Radio Galaxies (3CRR) at $1 < z < 2$. The bolometric luminosities are derived by integrating the intrinsic accretion disk spectral energy distributions (SEDs) over the range 1\,\micron--10\,keV (excluding reprocessed infrared emission) and integrating over inclination angles (to account for accretion disk emission anisotropy). We use accretion disk models, fitted to observed data, to self-consistently bridge the unobserved wavelength region between the UV and X-rays with no need for far-UV gap repair. The resulting bolometric corrections span a wide range
($\sim$1--400) across different wavelengths, showing minimal
dependence on redshift, X-ray luminosity, and  black hole mass,  which is possibly due to a narrow range of these intrinsic
AGN parameters in the sample. However, when the sample is divided by Eddington ratio, the X-ray bolometric corrections exhibit distinctly different values, with higher correction factors corresponding to higher Eddington ratios. We also examine the connection between
total radio luminosity and accretion disk power. For most 3CRR
sources, the radio power constitutes roughly 1\%--10\% of the
accretion disk luminosity. However, quasars with compact jets exhibit
higher radio luminosities for a given accretion disk power. This
suggests a higher efficiency of conversion of accretion power to radio
luminosity in the younger jets. Our
results provide physically motivated bolometric corrections for
powerful radio quasars that are applicable to powerful radio-loud quasars at any epoch.
\end{abstract}


\section{Introduction} \label{sec:intro}

Active Galactic Nuclei (AGN) release immense energy as they accrete matter, producing radiation that spans the full electromagnetic spectrum. Quantifying this total radiative output—the bolometric luminosity—is crucial for tracing the growth of supermassive black holes (SMBHs) and their impact on galaxies. Ideally, the bolometric luminosity would be measured by integrating the AGN’s spectral energy distribution (SED) over all wavelengths, yet observational coverage is typically incomplete. Therefore, empirical bolometric corrections are used instead to infer the total luminosity from observations within limited spectral ranges.  

The definition of what constitutes the total radiative output of an AGN remains an open issue \citep[for a full discussion, see][]{Azadi25}. Some studies define the bolometric luminosity as arising solely from the intrinsic emission of the accretion disk, typically integrating from the optical to the X-ray regime \citep[e.g.,][]{Marconi2004,Nemmen2010,R12,Azadi25}, while others also include the infrared contribution from dust heated by the central source \citep[e.g.,][]{Elvis1994,Richards2006}. The distinction reflects whether the reprocessed mid-infrared emission is considered part of the primary energy budget or a secondary manifestation of the same photons. In addition, anisotropy in the disk emission and relativistic effects—such as Doppler boosting, aberration, and light bending—impose a strong dependence on inclination angle, making it challenging to use uniform bolometric corrections for individual AGN. \citep[e.g.,][]{Hubeny2001,Nemmen2010}.

A central question is which wavelength band provides the most reliable proxy for bolometric luminosity, and which physical parameters—such as AGN luminosity, black hole mass, Eddington ratio, and/or spin—primarily govern it \citep[e.g.,][]{Marconi2004,V2007,Hopkins2007,Netzer2019,Azadi25}. In a recent study, \citet{Azadi25} investigated these questions by constructing a comprehensive grid of accretion-disk SEDs using the QSOSED model \citep{Kubota2018}. The integration was carried out over 1\,$\mu$m–10\,keV, encompassing emission from the accretion disk and corona as the primary radiative components while explicitly excluding reprocessed infrared radiation from the dusty torus. From this model grid, bolometric corrections were derived as a function of SMBH mass ($M_{\rm SMBH}$), Eddington ratio ($\lambda_{\rm Edd}$), spin($a\equiv c J /GM_{\rm SMBH}^{2}$), and inclination ($\theta$). The analysis provides a unified physical framework linking the intrinsic properties of the central engine to the observed luminosities and includes direct comparisons with earlier empirical and theoretical prescriptions. \citet{Azadi25} show that the bolometric output of an AGN is primarily determined by the accretion rate and SMBH mass, with smaller but noticeable effects from spin and inclination. Increasing SMBH mass produces cooler accretion disks with SEDs peaking at longer wavelengths, whereas higher Eddington ratios or larger spins yield hotter disks which SEDs peak in the far-UV. Inclination further modifies the observed SED, reducing the optical–UV emission (when inclination increases) while leaving the X-ray flux nearly isotropic. As a result, bolometric corrections in the optical–near-UV range ($\sim5100$–$3000$\,\AA) are most sensitive to SMBH mass, whereas X-ray corrections depend more strongly on the Eddington ratio. Near the SED peak in the far-UV ($\sim1450$\,\AA), these dependencies weaken, making this band a particularly robust and physically motivated reference for estimating bolometric luminosities \citep[see also][]{Azadi25}.

In this study, we apply the framework for deriving bolometric corrections based on accretion disk and corona emission developed by \citet{Azadi25} to a sample of 20 radio-loud quasars at $1<z<2$, drawn from the Revised Third Cambridge Catalogue of Radio Galaxies \citep[3CRR;][]{Laing1983}, in order to determine their intrinsic radiative power. 
The full multi-wavelength SEDs, spanning from radio to X-rays, and their detailed fits for this quasar sample were presented in \citet{Azadi2020}. We apply the bolometric correction methodology of \citet{Azadi25} to the best-fit accretion disk models from \citet{Azadi2020} to determine the bolometric luminosities and bolometric correction factors for the sample.  

The paper is organized as follows: Section~\ref{sec:rl} describes the sample and data; Section~\ref{sec:arxsed} summarizes the SED modeling presented in \citet{Azadi2020}; Section~\ref{sec:bc} presents the derived bolometric corrections; and Section~\ref{sec:summary} provides a summary of our results.

\section{Sample and Data}
\label{sec:rl}

In this study, we aim to determine the bolometric correction factors of the most powerful radio-loud quasars at $1<z<2$. 
For this purpose, we focus on sources from the Revised Third Cambridge Catalogue of Radio Galaxies \citep[3CRR;][]{Laing1983}. 
The full 3CRR catalog contains 173 FR~II radio galaxies out to $z<2.5$ and is 96\% complete down to a flux density of 10\,Jy at 178\,MHz. 
Within the redshift range $1<z<2$, the catalog includes 38 broad- and narrow-line radio galaxies \citep{Wilkes2013}, of which 20 are broad-line objects (i.e., quasars) studied in this work (see Table~\ref{tab:lbol}).  

We constructed radio-to-X-ray SEDs for these 20 quasars by combining new and archival photometry from multi-frequency radio observations, SMA, ALMA, \textit{Herschel}, \textit{WISE}, \textit{Spitzer}, 2MASS, UKIRT, SDSS, \textit{XMM-Newton}, and \textit{Chandra} \citep[see][for details]{Azadi2020}. 
This provides a complete, randomly-oriented sample with extensive wavelength coverage spanning nearly ten orders of magnitude in frequency.  The SEDs were modeled using the state-of-the-art \textsc{ARXSED} framework, as described below.

\begin{figure*}
\centering
    \includegraphics[height=0.7\linewidth,angle=90]{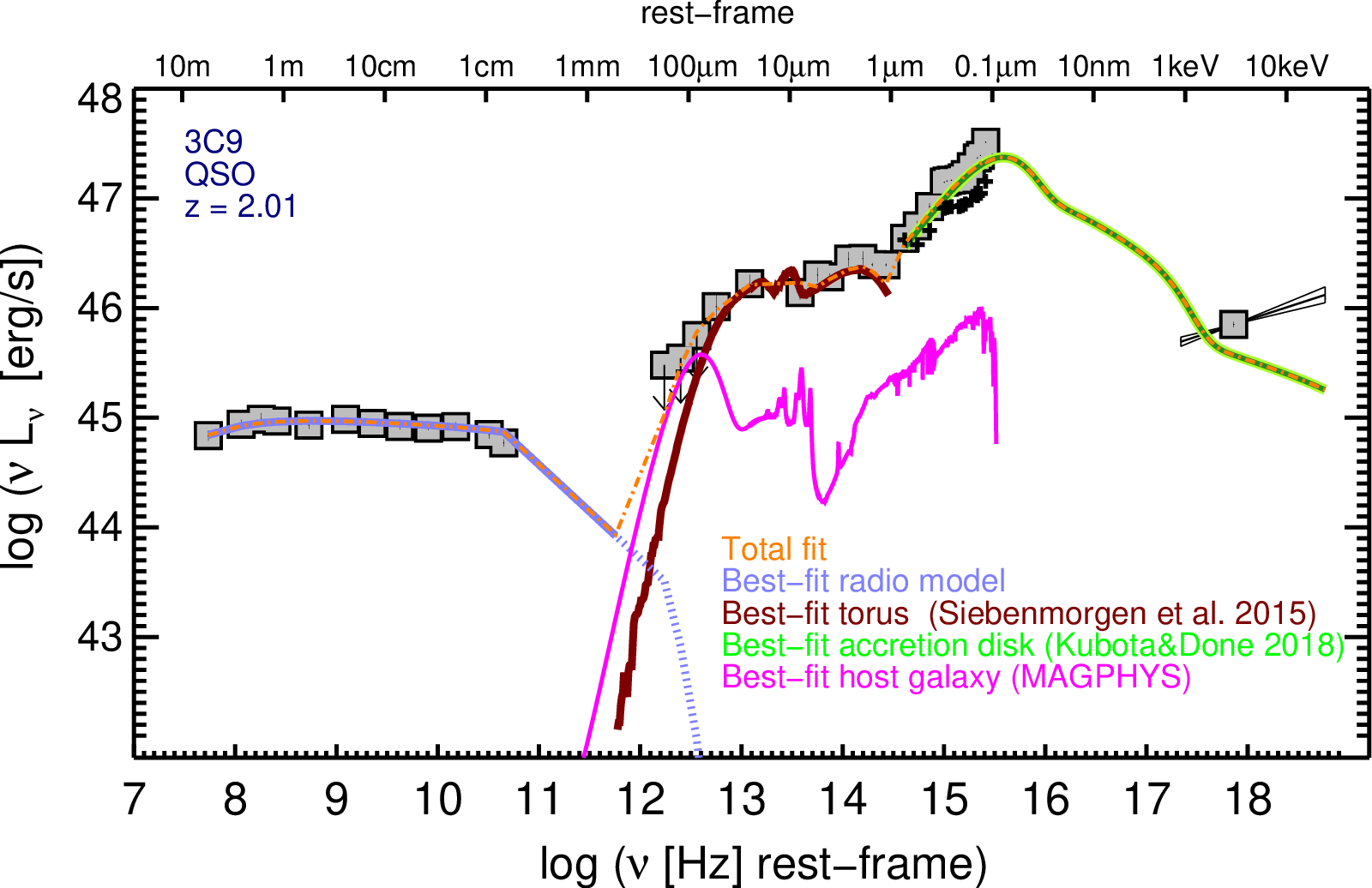}
    \caption{The intrinsic SED of 3C~9 from \citet{Azadi2020}. 
    Grey boxes show the photometry, with black plus signs indicating the observed photometric points prior to extinction correction. Colored lines trace the emission components of the best-fit ARXSED model. 
    AGN components include the accretion disk (green), torus (red), and radio emission (light blue). 
    The radio component (light blue) encompasses the lobes, and the dotted extension shows the synchrotron cutoff due to aging electron populations. The host galaxy contribution is shown in magenta.
    }    
    \label{fig:example}
\end{figure*}

\begin{table*}
\centering
\caption{3CRR quasars: redshift, intrinsic X-ray luminosity (0.3--8 keV), black-hole mass, Eddington ratio, and bolometric luminosity.}
\begin{tabular}{lccccc}
\hline\hline
Source & $z$ & $\log L_{0.3-8\,\mathrm{keV}}^{(1)}$ (erg s$^{-1}$) & $M_{\rm BH}^{(2)}$ ($10^{9} M_\odot$) & $\log\lambda_{\rm Edd}^{(3)}$ & $L_{\rm bol}$ (erg s$^{-1}$) \\
\hline
3C\,009   & 2.009 & 45.85 & 3.8 & $-0.4$ & $2.36\times10^{47}$ \\
3C\,014   & 1.469 & 45.97 & 9.6 & $-0.4$ & $5.25\times10^{47}$ \\
3C\,043   & 1.459 & 45.77 & 2.1 & $-1.0$ & $5.25\times10^{46}$ \\
3C\,181   & 1.382 & 45.71 & 1.7 & $-0.6$ & $6.71\times10^{46}$ \\
3C\,186   & 1.067 & 45.35 & 1.5 & $-0.6$ & $5.98\times10^{46}$ \\
3C\,190   & 1.195 & 45.54 & 6.0 & $-1.0$ & $1.04\times10^{47}$ \\
3C\,191   & 1.956 & 45.86 & 1.8 & $-0.6$ & $7.52\times10^{46}$ \\
3C\,204   & 1.112 & 45.86 & 2.9 & $-1.0$ & $5.58\times10^{46}$ \\
3C\,205   & 1.534 & 46.11 & 9.8 & $-0.8$ & $2.70\times10^{47}$ \\
3C\,208   & 1.110 & 45.65 & 1.5 & $-0.4$ & $1.06\times10^{47}$ \\
3C\,212   & 1.048 & 46.00 & 6.0 & $-1.2$ & $7.38\times10^{46}$ \\
3C\,245   & 1.029 & 45.84 & 1.1 & $-0.6$ & $5.33\times10^{46}$ \\
3C\,268.4 & 1.398 & 45.94 & 6.2 & $-1.0$ & $1.04\times10^{47}$ \\
3C\,270.1 & 1.532 & 45.87 & 4.6 & $-1.0$ & $7.01\times10^{46}$ \\
3C\,287   & 1.055 & 45.57 & 1.5 & $-1.2$ & $1.78\times10^{46}$ \\
3C\,318   & 1.574 & 45.50 & 2.4 & $-1.6$ & $1.14\times10^{46}$ \\
3C\,325   & 1.135 & 45.22 & 3.8 & $-1.4$ & $2.02\times10^{46}$ \\
4C\,16.49 & 1.880 & 46.03 & 2.1 & $-1.4$ & $1.61\times10^{46}$ \\
3C\,432   & 1.785 & 45.79 & 4.7 & $-1.2$ & $1.78\times10^{46}$ \\
3C\,454.0 & 1.757 & 45.70 & 0.8 & $\;\;0.0$ & $1.32\times10^{47}$ \\
\hline
\end{tabular}
\label{tab:lbol}
\begin{flushleft}
\footnotesize \textbf{Notes.}  
(1) $L_{\mathrm{X}}$ values are from \citet{Wilkes2013}.  
(2) $M_{\rm BH}$ values are from \citet{Azadi2020}.  
(3) Eddington ratios ($\lambda_{\rm Edd}$) are from \citet{Azadi2020}.
\end{flushleft}
\end{table*}

\section{ARXSED Model} \label{sec:arxsed}

\textsc{ARXSED} is a semi-empirical framework designed to reproduce the emission from all major AGN components simultaneously, including the radio lobes and jets, the dusty torus, the accretion disk, and the host galaxy. At radio wavelengths, the model accounts for emission from extended lobes, jets, cores, and hot spots using single or double power laws, parabolic functions, or combinations thereof. 
To capture the observed steepening of synchrotron spectra, \textsc{ARXSED} includes an exponential cutoff at high frequencies.  

In the infrared, \textsc{ARXSED} adopts the two-phase torus model of \citet{Siebenmorgen2015}, in which dust can reside in a homogeneous disk, a clumpy medium, or a mixture of both. 
At optical–UV–X-ray wavelengths, it incorporates the accretion-disk model of \citet{Kubota2018}, which divides the emission into three regions: 
(1) an inner hot corona with electron temperatures $kT_{e} \sim 40$--100\,keV producing hard Comptonization and hard X-rays; 
(2) an intermediate region with $kT_{e} \sim 0.1$--1\,keV producing soft Comptonization and soft X-rays; and 
(3) an outer region responsible for the thermal UV–optical continuum. We note that any non-thermal component associated with the radio structure is not included in the X-ray component (see Section \ref{sec:ad_vs_rad}).
Host-galaxy emission from radio to UV wavelengths is modeled using the \textsc{magphys} framework \citep{dc2008,dcr2015}. 
For the present sample, \textsc{ARXSED} yields intrinsic SEDs after correcting for reddening and absorption in the torus, the host galaxy, and the Milky Way.  

The modeling proceeds as follows. 
First, \textsc{ARXSED} corrects the photometry between 0.91 and 13\,\micron\ for Galactic absorption using the attenuation law $\tau_{\lambda} \propto \lambda^{-0.7}$ from \citet{charlot2000simple}. 
It then fits the torus emission and corrects the optical–UV photometry for internal obscuration \citep[see Equation~16 in][]{Azadi2020,Siebenmorgen2015}. 
X-ray luminosities are assumed to be intrinsic, corrected for both Galactic and intrinsic absorption \citep{Wilkes2013,Azadi2020}. 
To model the optical–UV–X-ray portion of the SED, \citet{Azadi2020} employed the \citet{Kubota2018} accretion-disk model, generating $\sim$11,000 templates spanning a large range of the main physical parameters: SMBH mass, Eddington ratio, spin, and viewing angle. 
Where available, emission-line features (e.g., \ion{Mg}{2} and \ion{C}{4} FWHM) were used as priors for SMBH mass, and templates were built accordingly. For sources without such spectroscopic information, the average values from those with prior mass estimates were used. However, this mass was used only as an initial estimate, allowing $M_{\rm BH}$ of the templates to vary within $\pm1$dex of that value.
The viewing angle of each accretion-disk template was restricted to within $\pm 12^{\circ}$ of the best-fit torus orientation, yielding a median inclination of $52^{\circ} \pm 5^{\circ}$ for this quasar sample.   

The best-fit radio-to-X-ray SEDs for the 20 quasars in our sample, along with estimates of the physical properties of the central engine (SMBH mass, Eddington ratio, spin), the torus (viewing angle, dust filling factor, optical depth), and the host galaxy (stellar mass, star formation rate), are presented in \citet{Azadi2020}. 
As an example, Figure~\ref{fig:example} shows the radio-to-X-ray SED and corresponding \textsc{ARXSED} fit for the quasar 3C~9.

\section{Estimating Bolometric Luminosity of 3CRR Quasars} 
\label{sec:bc}

To estimate the bolometric luminosities, and bolometric correction factors for each of the 3CRR quasars, we used the best-fitting accretion disk models from \textsc{ARXSED} \citep{Azadi2020}. 
The bolometric luminosity was obtained by integrating the intrinsic (rather than observed) SED from 1\,\micron\ to 10\,keV, and the bolometric correction factor is defined as

\begin{equation} \label{eq:bc_3crr}
    {\rm BC}(\nu,\theta) \equiv 
    \frac{\int_{0}^{\pi/2} \sin\theta \int_{1\,\mu{\rm m}}^{10\,\rm keV} L_{\nu}(\theta)\, d\nu\, d\theta}
    {\nu L_{\nu}(\theta)} ,
\end{equation}

for any specified frequency $\nu$ and inclination angle $\theta$. 
Although the \textsc{ARXSED} best fit corresponds to a single inclination angle, we integrate over all $\theta$ to fully account for the anisotropy of the accretion disk emission. 
Mid-infrared emission from the torus is excluded from the integration because it is dominated by reprocessed accretion-disk radiation already included in the model. 
The accretion-disk templates (see Figure~\ref{fig:example}) self-consistently bridge the otherwise unobserved region between the UV and soft X-ray bands, removing the need for empirical gap-repair methods. 
By integrating over all  angles, our approach naturally incorporates the geometric anisotropy of disk radiation, and thus no additional isotropy correction is required. 
Table~\ref{tab:lbol} lists the bolometric luminosities estimated for the 20 sources in our sample, calculated using Equation~\ref{eq:bc_3crr}.

\begin{figure*}[t!]
\centering
\includegraphics[width=0.7\textwidth]{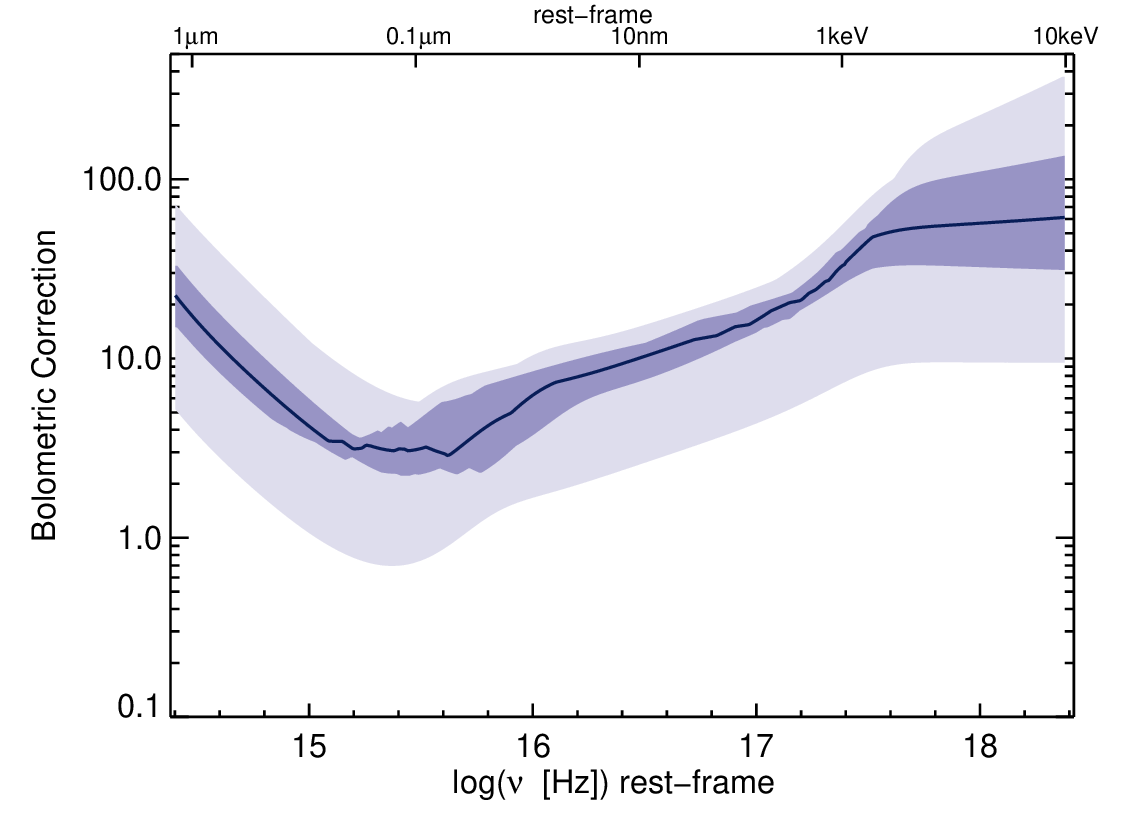}
\caption{
Bolometric correction factor as a function of frequency for the 20 3CRR radio-loud quasars at $z\sim1$ \citep{Azadi2020}. 
The solid curve represents the median bolometric correction. 
The dark purple shading indicates the interquartile (25th–75th percentile) range, while the light purple region shows the full range of values across the sample. 
Correction factors are derived from the best-fit  accretion disk models of \citet{Azadi2020}, using the integral defined in Equation~\ref{eq:bc_3crr}.
}
\label{fig:bc}
\end{figure*}

Figure~\ref{fig:bc} (tabulated in Table \ref{tab:bc_freq}) shows the bolometric correction as a function of rest-frame frequency for the 3CRR quasars at \z.
The light purple region indicates the full range of values across the sample, while the dark purple shading marks the interquartile range (25th–75th percentile), and the solid curve represents the sample median. 
The bolometric correction reaches a minimum near  $\log \nu \sim 15.4$, corresponding to a wavelength of $\sim$0.1\,\micron, where the accretion disk SEDs typically peak. 
At wavelengths farther from this peak, the bolometric correction increases, although the scatter within the interquartile range remains relatively small and nearly constant. 
Toward the X-ray regime, the percentile range in Figure~\ref{fig:bc} broadens, which primarily reflects the strong dependence of the bolometric correction in the X-ray range on the Eddington ratio \citep[see the discussion below and also ][]{Azadi25}. This behavior is intrinsically captured in the QSOSED model, as the accretion disk SEDs at X-rays are constructed to be independent of parameters such as  SMBH mass, spin, and inclination. The model assumes an increase in the Eddington ratio leads to an increase  in luminosity \citep{Ho2008} and is accompanied by a systematic steepening of both $\alpha_{\rm ox}$ and the X-ray photon index $\Gamma_{\rm 2-10\,keV}$ (see \citealt{Kubota2018} for a detailed discussion). The interplay of these trends results the bolometric correction in the X-ray band particularly sensitive to variations in the Eddington ratio.


\begin{table}
\centering
\caption{The median bolometric correction values (with their corresponding 25th--75th percentile ranges), plotted in Figure~\ref{fig:bc} as a function of frequency }
\begin{tabular}{cc}
\hline\hline
$\log \nu $ (rest-frame)& Bolometric Correction \\
\hline
14.5 & $16.1^{24.3}_{11.2}$ \\
14.7 & $7.86^{11.7}_{5.67}$ \\
15.0 & $4.17^{5.92}_{3.51}$ \\
15.3 & $3.26^{3.68}_{2.56}$ \\
15.5 & $3.14^{4.65}_{2.26}$ \\
15.7 & $3.77^{6.58}_{2.35}$ \\
16.0 & $6.26^{8.48}_{3.75}$ \\
16.3 & $8.21^{10.3}_{5.97}$ \\
16.5 & $10.3^{12.2}_{7.60}$ \\
16.7 & $12.9^{16.4}_{10.0}$ \\
17.0 & $16.4^{20.2}_{14.0}$ \\
17.3 & $23.6^{28.6}_{20.4}$ \\
17.5 & $45.1^{56.0}_{31.0}$ \\
18.0 & $56.9^{110.4}_{32.4}$ \\
\hline
\end{tabular}
\label{tab:bc_freq}
\end{table}

\subsection{Bolometric Correction Dependence on  Quasars' Properties}

\begin{figure*}[th!]
\begin{center}
\includegraphics[height=0.68\textwidth]{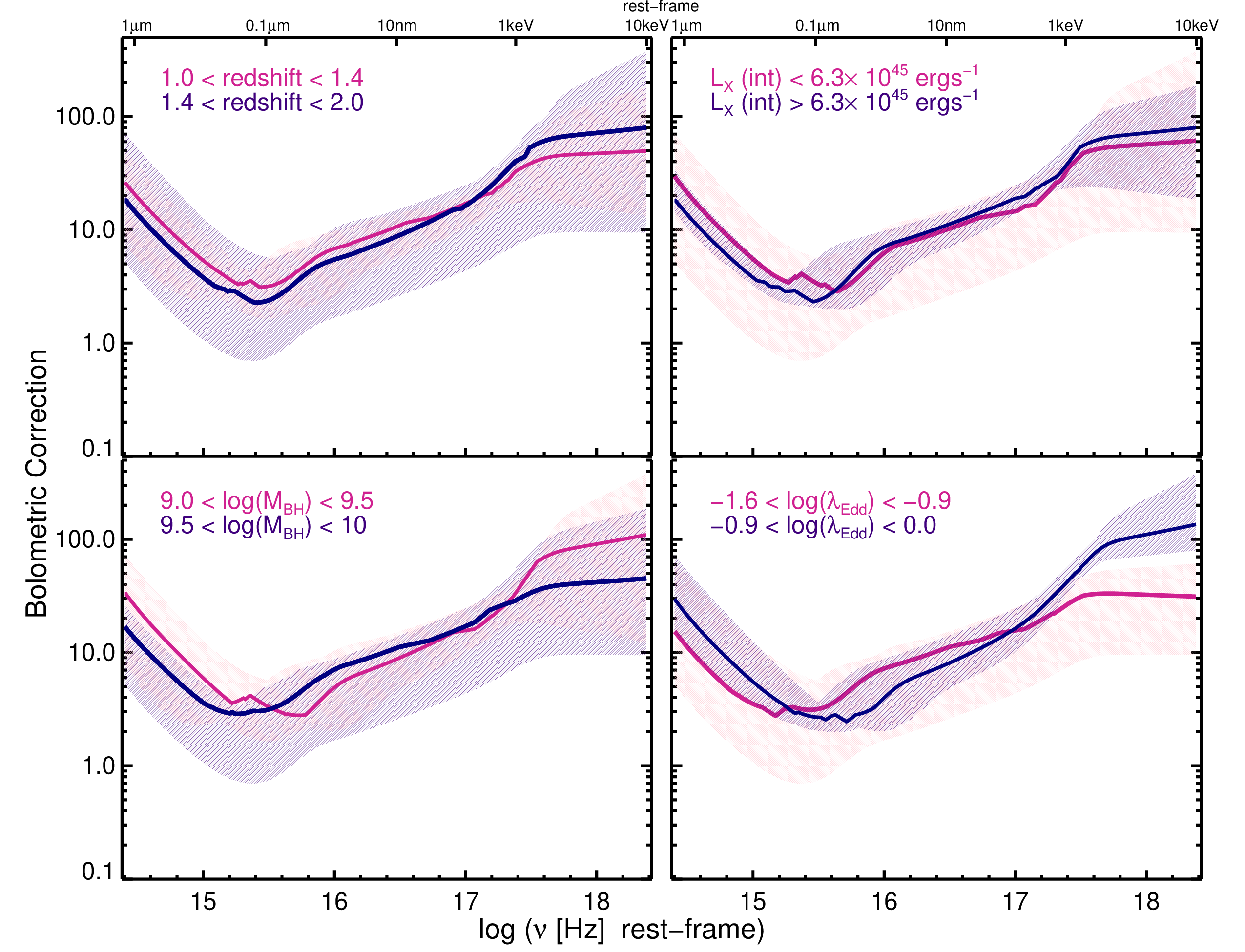}
\caption{
Bolometric correction factor as a function of frequency for the 3CRR quasars at \z. 
Each panel shows the sample divided into two bins based on the average value of a given parameter: redshift, X-ray luminosity, SMBH mass, and Eddington ratio. 
The intrinsic X-ray luminosities are adopted from \citet{Wilkes2013}, while the SMBH masses and Eddington ratios are derived from the \textsc{ARXSED} fits \citep[see Table~6 of][]{Azadi2020}. }

\label{fig:agn_prop}
\end{center}
\end{figure*}

Figure~\ref{fig:agn_prop} shows the bolometric corrections for the 3CRR quasars, divided into two bins based on the average value of each parameter in the sample: redshift, X-ray luminosity, SMBH mass, and Eddington ratio. 
The intrinsic X-ray luminosities are taken from \citet{Wilkes2013}, while the SMBH masses and Eddington ratios are derived from the \textsc{ARXSED} fits \citep[see Table~6 of][]{Azadi2020}. Because of the small sample size and the limited range of parameter values, the envelopes of the bolometric corrections largely overlap at most frequencies, indicating no significant dependence on these quantities. 

The redshift panel (top left) shows a broader range for the higher-redshift bin, although the frequency at which the bolometric correction reaches its minimum is nearly identical between the low- and high-redshift sub-samples (median curves at $\log \nu \sim 15.4$ for both). Both sub-samples exhibit a wider spread at X-ray frequencies, primarily due to variations in the Eddington ratio, as discussed below.

The top-right panel of Figure~\ref{fig:agn_prop} shows the sample divided according to X-ray luminosity, where $L_{\mathrm{X}}$ is the rest-frame $L_{\mathrm{0.3–8\,keV}}$ luminosity corrected for any significant intrinsic absorption, $N_{\mathrm{H,int}}$. 
Because all of our sources are quasars, they span only a narrow range in $N_{\mathrm{H,int}}$. 
For sources with low X-ray net counts, $N_{\mathrm{H}}$ was not well constrained, and a $3\sigma$ upper limit was adopted instead \citep[see][]{Wilkes2013}. 
In such cases, the intrinsic X-ray luminosity may be slightly overestimated. 
The bolometric correction reaches its minimum at slightly lower frequencies in the higher-$L_{\mathrm{X}}$ bin ($\log \nu \sim 15.5$) compared to the lower-$L_{\mathrm{X}}$ bin ($\log \nu \sim 15.6$). The low-$L_{\mathrm{X}}$ subsample exhibits a larger dispersion in bolometric correction, likely because of sources with lower X-ray counts and hence lower $L_{\mathrm{X}}$.

The trends with SMBH mass and Eddington ratio reflect how these parameters shape the accretion-disk SED \citep{Azadi25}. 
An increase in SMBH mass results in a cooler  accretion disk and shifts its SED peak toward lower frequencies, which in turn moves the minimum of the bolometric correction to smaller $\log \nu$. 
This trend is evident in Figure~\ref{fig:agn_prop}: the bolometric-correction minimum occurs at $\log \nu \simeq 15.3$ for the higher-mass bin, compared to $\log \nu \simeq 15.7$ for the lower-mass bin. 
At frequencies below this dip, more massive systems emit a larger fraction of their total luminosity, resulting in smaller bolometric corrections; at higher frequencies, the opposite holds true.


In contrast, variations in the Eddington ratio have the opposite effect. 
A higher Eddington ratio increases the accretion rate relative to the Eddington limit, producing a hotter disk whose emission peaks at higher frequencies in the far-UV. 
As a result, the minimum of the bolometric correction shifts toward higher $\log \nu$, occurring at $\log \nu \simeq 15.7$ for the higher-$\lambda_{\mathrm{Edd}}$ bin and at $\log \nu \simeq 15.2$ for the lower-$\lambda_{\mathrm{Edd}}$ bin. 
At frequencies above the SED peak, this trend reverses, as hotter disks radiate a larger fraction of their luminosity at higher energies. 
In the X-ray regime, however, accretion disk models with the same SMBH mass and spin converge to similar $\nu L_{\nu}$ values \citep[see][for details]{Azadi25}, resulting in the bolometric correction being insensitive to SMBH mass but increasingly dependent on the Eddington ratio.

\begin{table}
\centering
\caption{Bolometric correction values for the 3CRR quasars at five commonly used bands in the literature}
\begin{tabular}{cc}
\hline\hline
Wavelength (rest-frame) & Bolometric Correction \\
\hline
5100\,\AA & $7.38^{10.9}_{5.34}$ \\
3000\,\AA & $4.32^{6.17}_{3.59}$ \\
1450\,\AA & $3.13^{3.86}_{2.32}$ \\
2\,keV & $32.5^{40.2}_{25.7}$ \\
10\,keV & $61.3^{135.}_{31.2}$ \\
\hline
\end{tabular}
\label{tab:bc_wavelength}
\end{table}

\subsection{Advantages and Limitations of Determining the Bolometric Correction from Accretion Disk Modeling}
\label{sec:lim}

Estimating bolometric luminosities from the best-fit accretion disk model in \textsc{ARXSED} has several advantages as we elaborate below.

A persistent challenge in constructing AGN SEDs is the lack of coverage between the far-UV and X-ray regimes as it is inaccessible from ground- and space-based observatories, as discussed in detail by \citet{Azadi25}. 
Many previous studies have bridged this gap by interpolating in $\log\nu$--$\log(\nu L_{\nu})$ space \citep[e.g.,][]{Elvis1994,Richards2006,R12}. In contrast, \textsc{ARXSED} models the full SED self-consistently across the radio–to–X-ray range, predicting the spectral shape even in wavelength regions with limited observational data \citep[see][]{Azadi2020}.

The 3CRR quasars are among the most powerful AGN, typically powered by optically thick, geometrically thin accretion disks \citep[e.g.,][]{SS1976,NT1973}. 
These disks emit strongly across the visible–to–X-ray range, but their radiation is intrinsically anisotropic. 
Accurately estimating the total radiative output of such systems therefore requires accounting for inclination angle effects. 
Previous studies often simplified this problem by assuming a single, representative inclination angle for an entire population \citep[e.g.,][]{R12}, which can underestimate bolometric luminosities \citep[see discussion in][]{Azadi25}.  In this work, we avoid this approximation: the integration defined in Equation~\ref{eq:bc_3crr} explicitly includes the inclination dependence of the accretion disk emission, thereby incorporating anisotropic effects self-consistently without any additional correction.

Relativistic effects such as Doppler boosting, aberration, and light bending can substantially distort the observed SED of an AGN, leading to deviations from the expectations of a simple Newtonian disk \citep[e.g.,][]{Hubeny2001,Nemmen2010}. 
These effects become increasingly important at larger inclination angles, where relativistic aberration and beaming enhance the observed anisotropy of the emission \citep{Azadi25}.   
In \textsc{ARXSED}, the accretion disk emission is modeled using the QSOSED framework of \citet{Kubota2018}, which assumes that the disk truncates near the region responsible for hard X-ray production and does not include strong relativistic reflection or smearing effects \citep[e.g.,][]{Yaqoob2016,Porquet2018}.  Instead, the model incorporates a warm Comptonization component to reproduce the soft X-ray excess, consistent with recent evidence favoring Comptonization over relativistic reflection as the dominant origin of this feature \citep[e.g.,][]{Porquet2018}. 


The literature remains divided on the wavelength range that should be integrated to derive AGN bolometric luminosities (see Section~\ref{sec:intro}). 
Some studies argue that integrating only over the optical–to–soft X-ray regime provides the most physically meaningful estimate of the intrinsic emission \citep[e.g.,][]{Nemmen2010,R12,Azadi25}, while others include the mid-infrared contribution \citep[e.g.,][]{Richards2006}, suggesting that MIR emission, being more isotropic, may better trace the total radiative power. 
Our approach resolves this ambiguity: \textsc{ARXSED} first models the dusty torus and corrects the optical–UV photometry for obscuration accordingly. 
Integration over 1\,\micron--10\,keV thus captures all primary emission from the accretion disk- the same radiation that powers the reprocessed MIR emission—without double counting.

Another advantage of this approach is that it isolates the radiative power from the accretion disk and corona, excluding X-ray emission associated with radio structures such as jets or compact cores (see SED example in Figure~\ref{fig:example}). In sources with strong radio cores, empirical bolometric estimates based directly on observed X-ray fluxes may therefore overpredict the intrinsic disk radiative output by including non-coronal emission.

\begin{figure*}[ht!]
\centering
\includegraphics[width=0.6\textwidth,]{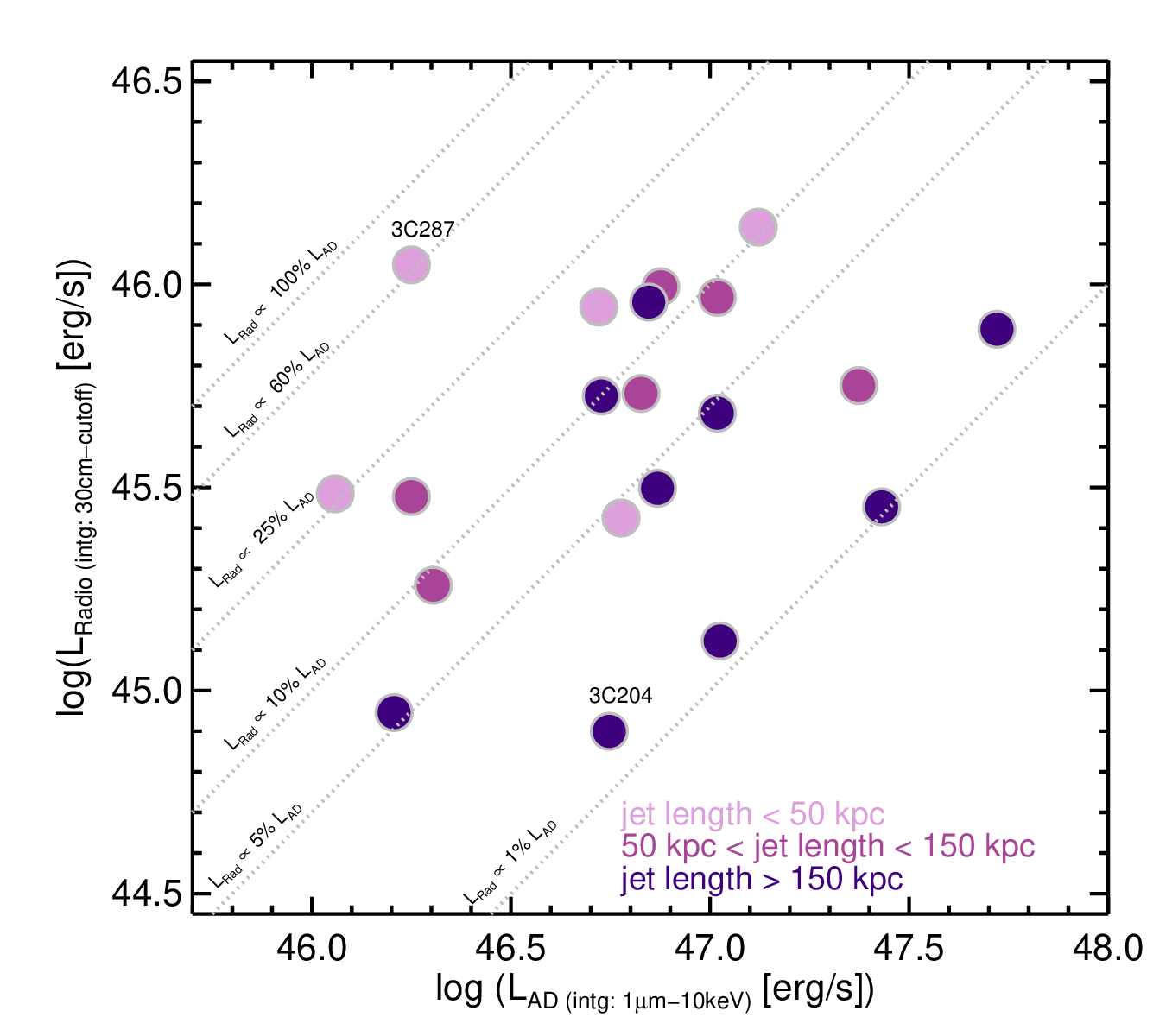}
\caption{The integrated radio luminosity is plotted against the integrated accretion disk luminosity for quasars hosting young/compact (light purple), intermediate (medium purple), and mature/extended (dark purple) jets. The dotted lines indicate different ratios between the integrated radio and accretion disk luminosities, ranging from 1\% to 100\%. The SED of the two extreme cases—3C 204 (lowest integrated radio luminosity) and 3C 287 (highest integrated radio luminosity)—are shown in Figure~\ref{fig:extreme}.}
\label{fig:rad_vs_ad}

    \centering
   \includegraphics[width=0.999\linewidth,clip,trim=10 220 100 360]{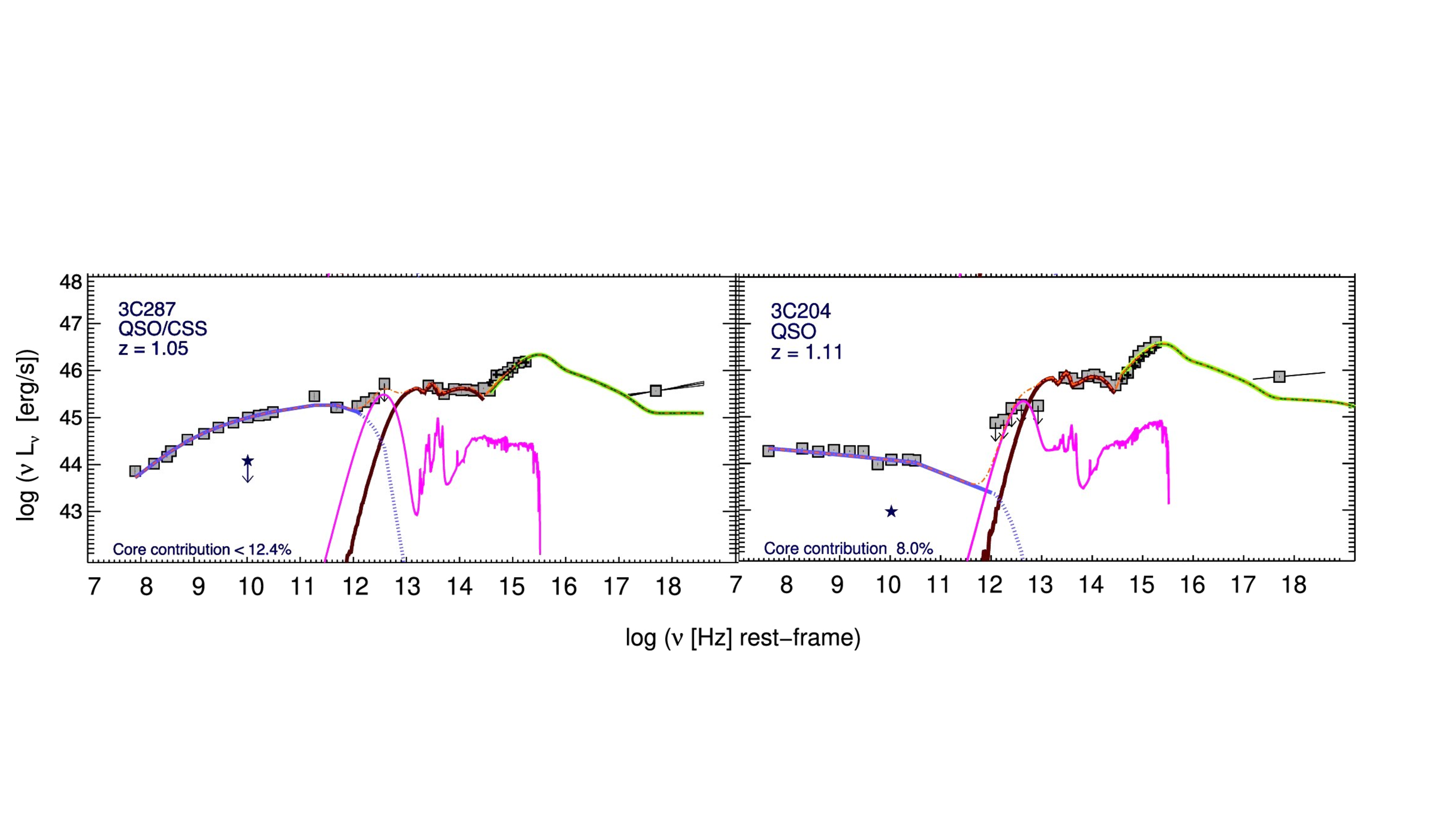}
    \caption{The SED of the two quasars with the highest and lowest integrated radio luminosities, adapted from \citet{Azadi2020}. The gray points show the broadband photometry corrected for absorption, while the black plus indicate the visible–UV photometry prior to applying the torus obscuration correction. The colored curves highlight the key emission components: light blue represents the radio emission from the core, jets, hot spots, and lobes (included in the radio integration); dark red corresponds to the infrared radiation from the torus; green traces the thermal emission from the accretion disk extending from the visible to X-ray regime; magenta denotes the stellar emission from the host galaxy; and orange shows the overall combined fit. The dotted light-blue section marks the cutoff frequency where the radio spectrum declines. The dark-blue star indicates the estimated core contribution to the 5 GHz flux density. At X-ray, the black line (or bow tie) represents the power-law model fitted to the absorption-corrected data.}  
    \label{fig:extreme}
\end{figure*}

Despite these advantages, one limitation of our estimates is the lack of reliable optical or UV photometry for a few objects (e.g., 3C\,325 and 4C\,16.49), which leads to larger uncertainties in the accretion disk modeling and less tightly constrained parameters.

Table~\ref{tab:bc_wavelength} lists the median bolometric correction values derived for the 3CRR quasars at $1<z<2$, at bands commonly used in the literature. Our results are consistent with, and fall within the range of, previously reported values \citep[e.g.,][]{Elvis1994,Richards2006,R12}. However, those reference samples are largely mixed and predominantly composed of radio-quiet quasars \citep[for details, see][]{Azadi25}. In contrast, the values reported here are specifically applicable to the radio-loud population. Furthermore, because the QSOSED model is not explicitly redshift-dependent, these bolometric corrections can be robustly applied to radio-loud AGN across cosmic time.

\subsection{The Integrated Power at Radio Frequencies} \label{sec:ad_vs_rad}

Our sample of 3CRR quasars at \z\ represents the most luminous radio sources at this epoch. 
Because the 3CRR catalog is radio-flux-limited, the sample is biased toward quasars with high radio and accretion disk luminosities. 
Figure~\ref{fig:rad_vs_ad} illustrates the relationship between the total radio luminosity and the integrated accretion disk (bolometric) luminosity for these quasars. 
The radio luminosities were computed from the best-fit radio models, integrated from $10^{8}$\,Hz up to the cutoff frequency (see Figure \ref{fig:extreme}, and model parameters in Table~7 of \citealt{Azadi2020}). This cutoff marks the frequency at which energy losses in the electron population cause a rapid decline in radio emission \citep[e.g.,][]{Blandford1979,Konigl1981}. In our sample it ranges between $10^{11}$ and $10^{13}$\,Hz, being higher in quasars with strong core emission \citep{Azadi2020}. 
The integrated radio luminosities therefore include contributions from all radio components—lobes, jets, cores, and hot spots.

In Figure~\ref{fig:rad_vs_ad}, the sample is divided into three groups based on the de-projected jet length, which also serves as a proxy for jet maturity, with the shortest jets typically representing the youngest systems \citep{Azadi2020}. Figure~\ref{fig:rad_vs_ad} shows that quasars with smaller jets generally exhibit higher radio luminosities relative to their accretion disk luminosities than those extremely extended jets ($>150$ kpc), suggesting that compact radio quasars may be more efficient at converting accretion power into radio emission, whereas the radio luminosity declines as the jets expand and age.
In the majority of the quasars with medium or extended jets, the radio power constitutes approximately 1--10\% of the accretion-disk power. The only compact source within this range is 3C 186, whose radio images and SED indicate that it is likely embedded in a dense ISM that hinders the expansion of its radio structure \citep{Azadi2020}. The source with the lowest integrated radio luminosity is 3C~204, whose SED is shown in Figure~\ref{fig:extreme}.

Sources above the 10\%  line tend to have bright cores and/or compact radio jets. This is especially evident for 3C\,287, a compact steep-spectrum (CSS) source with an 8\,kpc de-projected jet length and a radio SED that remains core-dominated out to infrared wavelengths (see Figure \ref{fig:extreme} adopted from \citealt{Azadi2020}).

Figure \ref{fig:extreme} shows that in our sources  an additional non-thermal X-ray component, likely arising from synchrotron self-Compton (SSC) or inverse Compton (IC) processes is required. Such emission may affect both the torus and accretion-disk fits, and consequently the derived luminosities. 
If this additional X-ray emission were included, the total radiative power of the radio structures would further increase, likely enhancing the observed separation between compact and extended sources. We also note that the integrated radio luminosity combines emission from multiple components—lobes, jets, cores, and hot spots—representing activity over different epochs of an AGN’s evolution. In contrast, the accretion disk luminosity traces the current accretion state. Thus, Figure~\ref{fig:rad_vs_ad} effectively compares radiative power over different timescales. While our findings suggest that compact quasars are brighter radio emitters than their extended counterparts, confirming the nature of this relation will require a larger and more statistically complete sample.

\section{Summary} \label{sec:summary}

In this study, we applied the methodology developed by \citet{Azadi25} to a sample of 20 radio-loud quasars from the 3CRR catalog at \z\ \citep{Azadi2020} for determing their bolometric luminosity. For each source, we determined the bolometric luminosity by integrating the best-fit accretion disk model over the range 1\,\micron--10\,keV and viewing angles from $0^\circ$ to $90^\circ$. 
Our main findings are summarized below:

\begin{itemize}

\item The bolometric correction factors in the 3CRR quasar sample span a wide range—from $\sim$1 to $\sim$400 across different wavelengths—although the interquartile range (25th-75th) shows considerably less variation. 

\item Our bolometric corrections at commonly used wavelengths (5100\,\AA, 3000\,\AA, 1450\,\AA) and X-ray energies (2\,keV and 10\,keV) are consistent with values reported in the literature, although they exhibit a broader range in the X-ray bands, reflecting the dependence of the X-ray bolometric correction on the Eddington ratio.

\item We find no significant dependence of the bolometric correction on redshift, X-ray luminosity, or black hole mass among the 3CRR quasars. Because these quasars occupy a relatively narrow range of intrinsic properties, their overall range of corrections overlaps substantially (Figure~\ref{fig:agn_prop}). However, when the sample is divided by Eddington ratio, the two sub-samples show clearly distinct distributions at X-ray energies—for example, at 10\,keV the correction increases from about 30 in the low-Eddington group to roughly 105 in the high-Eddington group, highlighting the dominant impact of Eddington ratio on the X-ray bolometric correction.

\item We find no strong correlation between total radio and accretion disk luminosities (Figure~\ref{fig:rad_vs_ad}). For most quasars, the radio power constitutes approximately 1--10\% of $L_{\rm AD}$. Nevertheless, the more compact radio sources tend to be brighter in radio emission at a given $L_{\rm AD}$, suggesting that compact radio quasars may be more efficient at converting accretion power into radio luminosity, while the radio output declines as the jets expand and age. Larger samples will be required to confirm these trends with higher statistical significance.

\end{itemize}

\section*{Acknowledgements} 

Support for this work was provided by NASA grants: \#80NSSC18K1609,  \#80NSSC19K1311,
 \#80NSSC20K0043  (MAz), and \#80NSSC21K0058 (JK), and by NASA Contract NAS8-03060   \textit{Chandra} X-ray Center (CXC), which is operated by the Smithsonian Astrophysical Observatory   
(BJW, JK). BJW acknowledges the support of the Royal Society 
and the Wolfson Foundation.
The scientific results in this article are based to a significant degree on observations made by the \textit{Chandra}~X-ray Observatory (CXO).
This research has made use of data obtained from the \textit{Chandra} Data
Archive. This research is based on observations made by {\it Herschel}, which
is an ESA space observatory with science instruments provided by European-led Principal Investigator consortia and with important
participation from NASA. This work is based in part on observations made with the Spitzer Space
Telescope, which was operated by the Jet Propulsion Laboratory, California Institute of Technology under a contract with NASA.

We acknowledge the use of Ned Wright's calculator
\citep{2006PASP..118.1711W} and NASA/IPAC Extragalactic Database (NED), operated by the Jet Propulsion Laboratory, California Institute
of Technology, under contract with the National Aeronautics and Space
Administration.

The authors would like to thank Chris Done, Ma\l gosia Sobolewska, Mark Birkinshaw and Diana Worrall for helpful comments that improved the quality of the paper.

\bibliographystyle{apjurl}
\bibliography{references.bib}
\end{document}